\documentclass[a4paper]{article}
\usepackage{graphicx}
\begin{document}

\begin{center}
\bf\Large Dynamics of  Anti-Proton -- Protons and Anti-Proton -- Nucleus Reactions
\end{center}

\begin{center}
(Reported at International Workshop on Nuclear Theory IWNT-35, Rila 2016, Bulgaria) 
\end{center} 

\begin{center}
A. Galoyan\footnote{VBLHEP, JINR, Dubna 141980,  Russia},
A. Ribon\footnote{CERN, CH-1211 Geneva 23, Switzerland},
V. Uzhinsky\footnote{LIT, JINR, Dubna 141980, Russia}
\end{center}

\begin{center}
\begin{minipage}{14cm}
A short review of simulation results of anti-proton-proton and
anti-proton-nucleus interactions obtained with Geant4 FTF
(Fritiof) model is presented. The model uses the main
assumptions of the Quark-Gluon-String Model or Dual Parton
Model. The model assumes production and fragmentation of
quark-anti-quark and diquark-anti-diquark strings in the
mentioned interactions. Key ingredients of the model are cross
sections of string creation processes and an usage of the LUND
string fragmentation algorithm. 
The determined  cross sections and improvements of the algorithm 
allow to satisfactory describe
a large set of experimental data, in particular, strange particle production,
$\Lambda$ hyperons and $K$ mesons.
\end{minipage}
\end{center}

\section{Introduction}
Simulations of anti-proton and anti-nucleus interactions with
protons and nuclei are needed for cosmic space experiments like
PAMELA, BESS, AMS, CAPRISE and so on which are going to search
for anti-nuclei in cosmic rays. They hope this will bring a new
light on the question about anti-matter in our Universe. A
detailed experimental study of anti-proton-proton and
anti-proton-nucleus interactions is foreseen at the future FAIR
facilities (GSI, Darmstadt, Germany) by the PANDA
Collaboration. The accelerator experiments and the cosmic ray
experiments need estimations of properties of the mentioned
interactions ($\bar pA$, $\bar AA$) for reconstruction and
identification of anti-nuclei and reaction products. This
requires a computer code for a good simulation of the
reactions. Recently, it becomes possible to simulate the
interactions in the well-known Geant4 toolkit \cite{Geant4}. At
a creation of the simulation code, we used the main assumptions
of the Dual-Parton or Quark-Gluon-String model (DPM/QGSM)
\cite{DPM}. For calculations of anti-proton-nucleus cross
sections, we used the Glauber approach
\cite{Glauber,FrancoAA,CzyzAA}. An extension of DPM/QGSM for
anti-proton-nucleus reactions was proposed also. The main
simulation results and assumptions of the model well be
presented below.

\section{Main processes of $\bar pp$ interactions}
\begin{figure}[cbth]
\begin{center}
\includegraphics[width=100mm,height=75mm,clip]{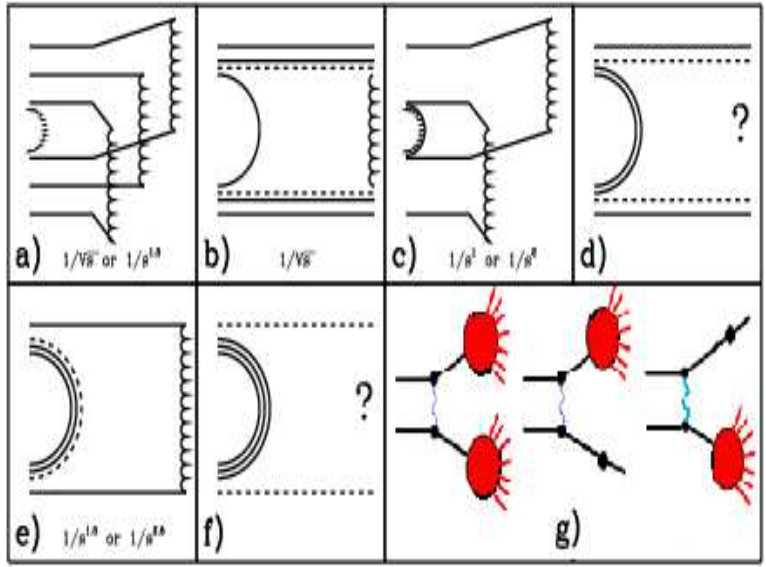}
\end{center}
\caption{Quark flow diagrams of $\bar pp$ interactions.}
\label{Fig1}
\end{figure}

Usually, the following list of diagrams shown in Fig.~\ref{Fig1} is considered in high energy phenomenology
for anti-proton-proton interactions.
The diagram "a" corresponds to string junction's annihilation with 3 quark-anti-quark string's creation.
The string junction is a gluonic object which couples together quarks in baryon. It is presented by
dashed line in the figure.
The diagram "b" represents quark and anti-quark annihilation leading to diquark-antidiquark string
creation. The diagram "c" corresponds to quark-anti-quark and string junction's annihilations with
creation of 2 quark-anti-quark strings. Diagrams "d" and "f" can be responsible for exotic meson production.
The last diagrams at the bottom of the figure are important at high energies. They are connected with
pomeron exchange in t-channel.

Energy dependencies of the cross sections of the processes in some cases are predicted by
the reggeon phenomenology. They are also shown in the figure. An elaborated scheme for a calculation
of the cross sections in the phenomenology was proposed in our papers \cite{Xs,Leap05}. As was shown
in the papers, the approach is valid at $P_{lab}\ >$ 3 GeV/c. Because experimental studies require much
less energies, we undertook a new attempt to estimate the cross sections. It is presented in the paper
\cite{Baldin}.
\begin{equation}
\sigma_a=25/\sqrt{s-4m^2} \ \ \ (mb),
\end{equation}
\begin{equation}
\sigma_b=15.65+700\cdot (2.172-\sqrt{s})^{2.5}\ \ \ (mb), \ \ \ \sqrt{s}\leq 2.172\ GeV,
\end{equation}
\begin{equation}
\sigma_b=34/\sqrt{s}\ \ \ (mb), \ \ \ \sqrt{s}> 2.172\ GeV,
\end{equation}
\begin{equation}
\sigma_c=\frac{2}{\sqrt{s=4m^2}}\left(\frac{m_p+m_t}{s}\right)^2 \ \ \ (mb),
\end{equation}
\begin{equation}
\sigma_e=140/s\ \ \ (mb),
\end{equation}
Processes without annihilations dominate at high energies. They
are presented in the bottom of Fig.~\ref{Fig1}. Their cross
sections can be described by
\begin{equation}
\sigma_{FTF}=35\cdot(1-2.1/\sqrt{s})\ \ \ (mb).
\end{equation}
For a fragmentation of the strings, we use the standard LUND
fragmentation algorithm implemented in Geant4.

\begin{figure}[cbth]
\begin{center}
\includegraphics[width=100mm,height=65mm,clip]{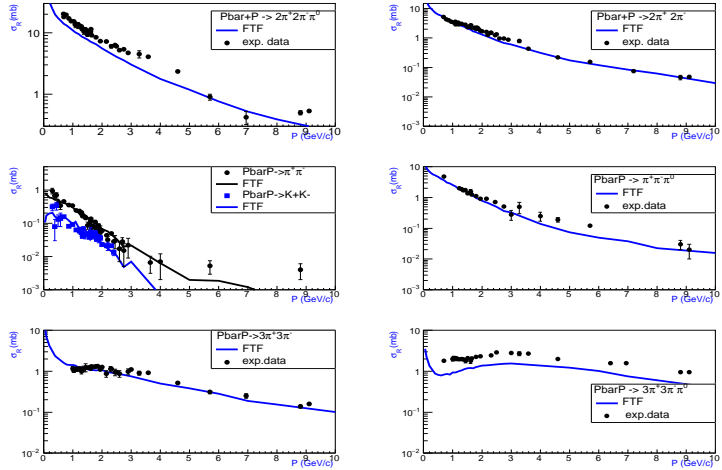}
\end{center}
\caption{Cross sections of $\bar pp$ interactions 
with only mesons in final states. The points are
experimental data. The lines are FTF model calculations.}
\label{Fig2}
\end{figure}

\begin{figure}[cbth]
\begin{center}
\includegraphics[width=100mm,height=65mm,clip]{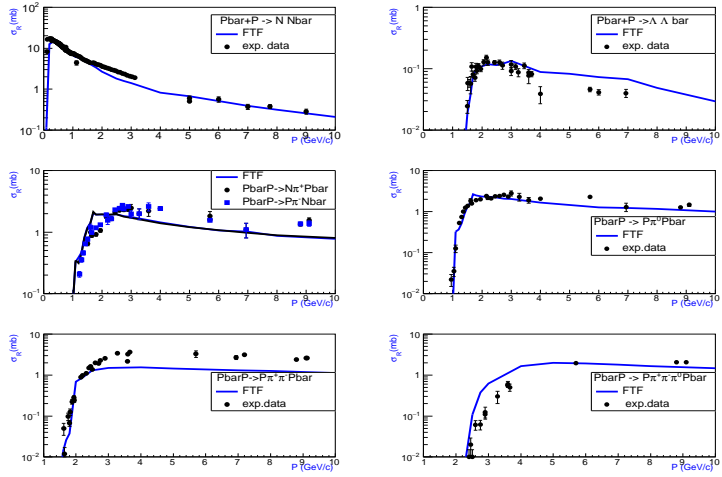}
\end{center}
\caption{Cross sections of $\bar pp$ interactions   
with baryons in final states. The points are
experimental data. The lines are FTF model calculations.}
\label{Fig3}
\end{figure}

All of these allows to simulate the main channels of $\bar pp$
interactions (see Figs. \ref{Fig2} and \ref{Fig3}).

As seen, the estimated cross sections, the improved string
fragmentation algorithm and the FTF model of Geant4 allow to
describe general properties of $\bar pp$ interactions. Thus, we
conclude that the existing phenomenology of anti-baryon-baryon
annihilation and hadron-nucleon high energy reactions with
corresponding cross sections can be considered as a unified
base for understanding of the $\bar pp$ interactions.

\section{$\bar pA$ interactions}
The first task in a simulation of the interactions is an
estimation of anti-proton-nucleus and anti-nucleus-nucleus
interaction cross sections (total, elastic and inelastic). It
is natural to use the Glauber approach for calculations of the
cross sections. For the first time, a good description of
elastic anti-proton-deuteron scattering was reached in the
classical paper by V.~Franco and R.J.~Glauber \cite{Franco66}
in 1966. After that, in 1985 O.D.~Dalkarov and V.A.~Karmanov
\cite{Dalk_Karm} showed that elastic and inelastic (with
excitation of nuclear levels) anti-proton scattering on C, Ca,
and Pb nuclei are described quite well at $\bar p$ kinetic
energies of 46.8 and 179.7 MeV within the approximation. The
Glauber approach in the question was also used in many other
papers. We applied the approach in our paper \cite{OurPaperPL}
and obtained the results presented in Fig. 4. As seen, the
agreement between the experimental data and the calculations is
rather good. More details of the calculations and results are
presented in \cite{OurPaperPL}. A code for calculation of the
total, elastic and inelastic cross sections of  anti-protons
and light anti-nuclei interactions with nuclei based on the
Glauber approach is implemented  in Geant4 toolkit. This code
can be activated using  PhysicsList FTFP\_BERT.

\begin{figure}[cbth]
\begin{center}
\includegraphics[width=20pc,height=45mm]{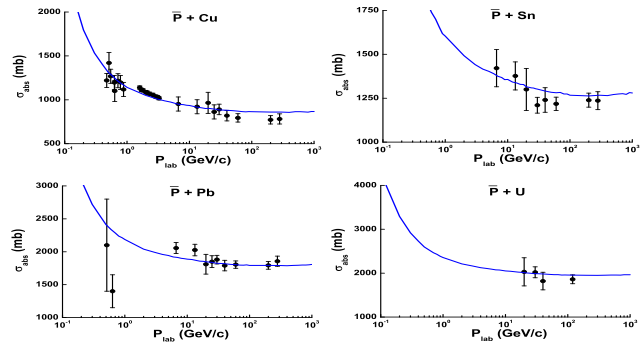}\hspace{2pc}
\end{center}
\caption{Absorption cross sections of anti-proton interactions with nuclei.
The points are experimental data (see references in \protect{\cite{OurPaperPL}}), 
the lines are our  calculations.}
\label{Fig4}
\end{figure}

It is known that a projectile can have multiple intra-nuclear
collisions in a target at high energies, or involves some
nuclear nucleons at the fast stage of interactions.
Distribution on the multiplicity of the involved nucleon,
$\nu$, can be found applying the asymptotic
Abramovsky-Gribov-Kancheli (AGK) cutting rules \cite{AGK} to
elastic scattering amplitude. For applications of the rules at
low energies, we proposed \cite{AGaloyanSup} "finite energy
corrections" to the rules, which decrease $\nu$ at low
energies. These allowed to extend the FTF model to low energy
domain.

In the first rough approximation, one can assume that an
interaction of an anti-baryon with a nuclear nucleon is
identical to the interaction with a free nucleon neglecting the
binding energy of the nucleon, $\sim$ 10 MeV. Though,
multiplicity of particles in hadron-nucleus interactions is
larger than in hadron-nucleon ones. This is explained by the
secondary particles cascading within a nucleus. The existing
intra-nuclear cascade models have passed a long history of
development and allow to simulate  meson and baryon
interactions with nuclei satisfactorily. Two of such models are
present in Geant4 -- the Bertini-like cascade model (BERT) and
the binary cascade model (BIC) \cite{BIC}. There is also a
simplified model -- the precompound model interface (PRECO),
which only absorbs low energy particles ($E\leq$ 10 MeV)
produced by a high energy generator and located in a nuclear
residual and ascribes them to the residual nuclei. The binary
cascade model and the precompound model can be easily coupled
with the FTF model. Therefore, we use in our following
calculations two combinations of the models: FTF+BIC and
FTF+PRECO\footnote{These combinations are denoted as FTFB and
FTFP, respectively.}. FTFP demonstrates properties of
interactions without intra-nuclear cascading. FTFB shows
effects of the all cascading processes.

As seen in Fig.~5, FTFB describes the general properties of
anti-proton-nucleus interactions, qualitatively. FTFP gives low
yields of protons and neutrons. Thus, accounting of the
particle cascading is very important for a correct simulation
of the reactions.

\begin{figure}[cbth]
\begin{center}
\includegraphics[width=75mm,height=60mm,clip]{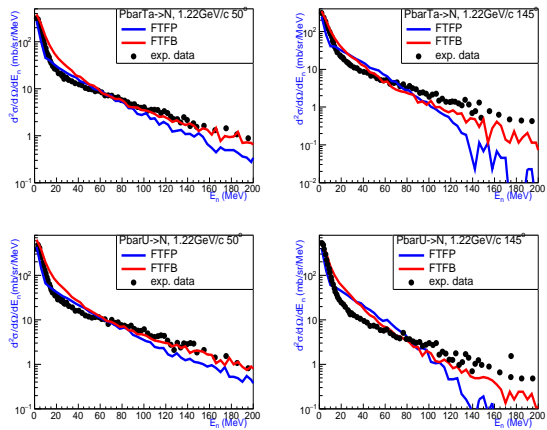}
\end{center}
  \caption{Neutron distributions on kinematical variables in anti-proton-nucleus
           interactions at $P_{lab}$=1.22 GeV/c. The points are experimental data \protect\cite{PbarAlow}. The lines
           are our calculations. The red and blue lines are calculations
           with FTFB and FTFP, respectively.}
\label{Fig5}
\end{figure}

\section{Strange particle production}
One of the aims of the PANDA experiment is a study of
properties of hyper nuclei. It is planned that hyperons will be
produced at interactions of anti-proton with hydrogen or
nuclear target and will be absorbed by another nucleus forming
a hyper nucleus. To reach the aim, it is needed to estimate
multiplicity of hyperons and correctly reproduce their
kinematical characteristics. Thus, we calculated with the FTF
model average multiplicities of $\Lambda$-hyperons and $K^0_S$-
mesons in anti-proton-proton interactions and presented them in
Fig. 6. As seen, the model correctly reproduces the
multiplicities below 10 GeV/c. At higher energies, the
experimental data scatter too strongly, and it is complicated
to draw a solid conclusion. The FTF model can also predict
production of $\Sigma^{\pm\ 0}-$ and $\Xi^{0-}-$ hyperons, as
well as their anti-particles.

\begin{figure}[cbth]
\begin{center}
\includegraphics[width=110mm,height=40mm,clip]{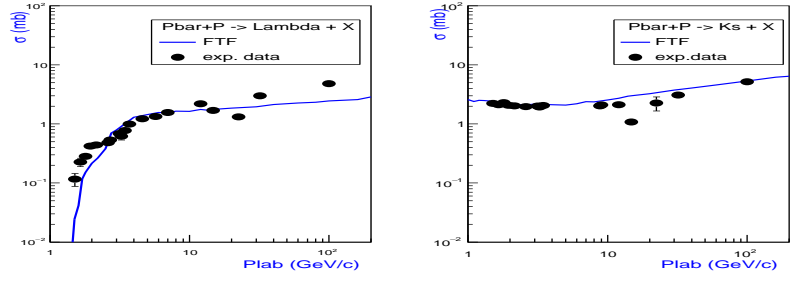}
  \caption{Average multiplicities of $\Lambda$-hyperons and $K^0_S$ mesons as functions of projectile
 momentum. The points are experimental data, the lines are the FTF model calculations.}
\end{center}
\label{Fig6}
\end{figure}

$\Lambda$-hyperons are mainly produced in the process of
Fig.~1b at a creation of $s\bar s$ quark's pair in a
diquark-anti-diquark string. The analogous pair production
takes place in the process of Fig.~1e.
 It is responsible for creation of $K^+K^-$ and $K^0K^0$ pairs at low energies.
  At high energies, the bottom processes
of Fig.~1 start to dominate, which lead to slow increase of the
strange particle multiplicities. Distributions of the strange
particles on kinematical variables are very important for
experiments. Calculated rapidity distributions of
$\Lambda$-hyperons and $K^0_S$ mesons in $\bar pp$ interactions
at momenta 3.6, 12 and 100 GeV/c  are presented in Fig. 7. As
seen, there is a reasonable agreement with the experimental
data. \vspace{-5mm}
\begin{figure}[cbth]
\begin{center}
\includegraphics[width=95mm,height=35mm,clip]{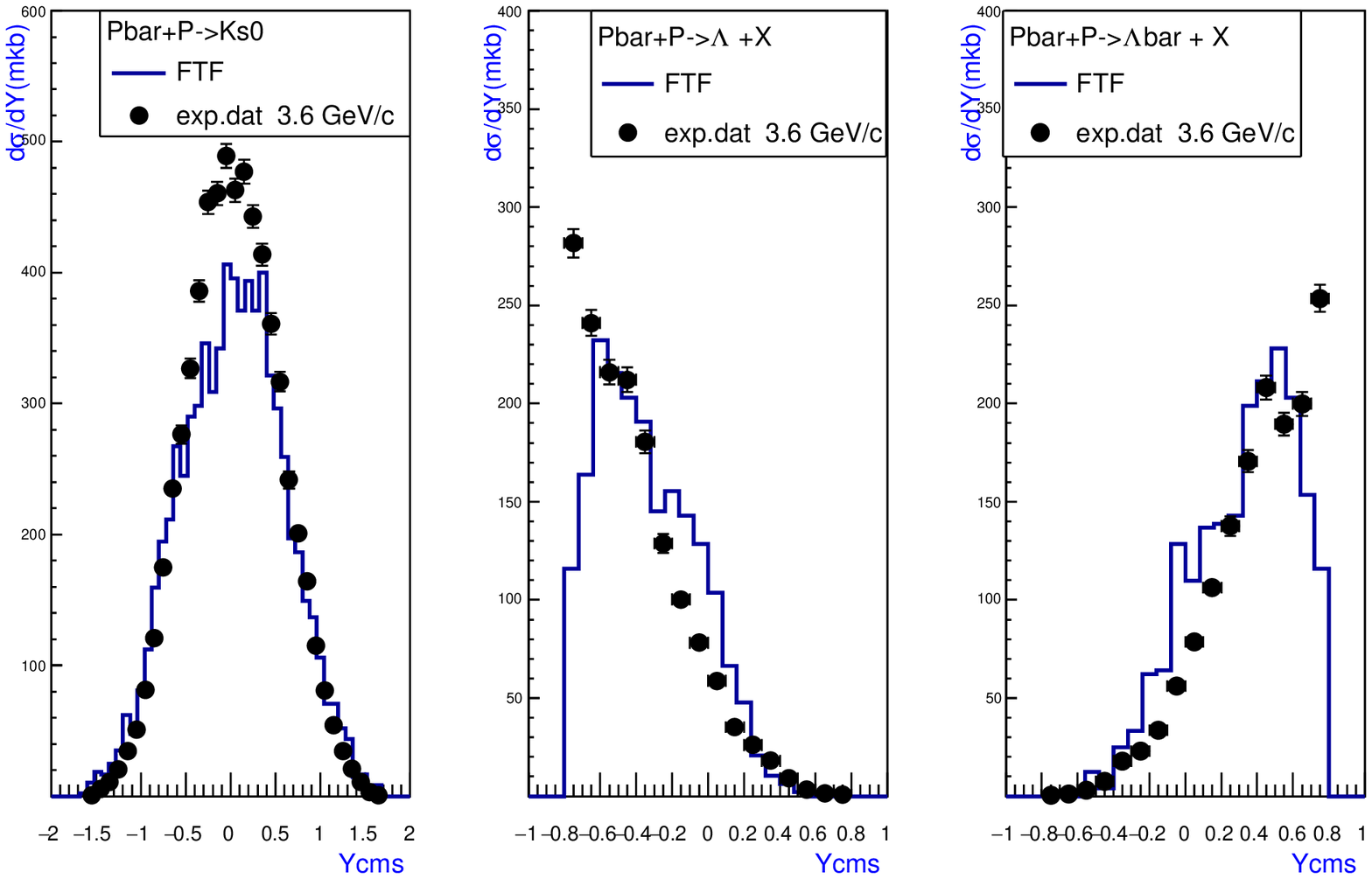}
\includegraphics[width=95mm,height=35mm,clip]{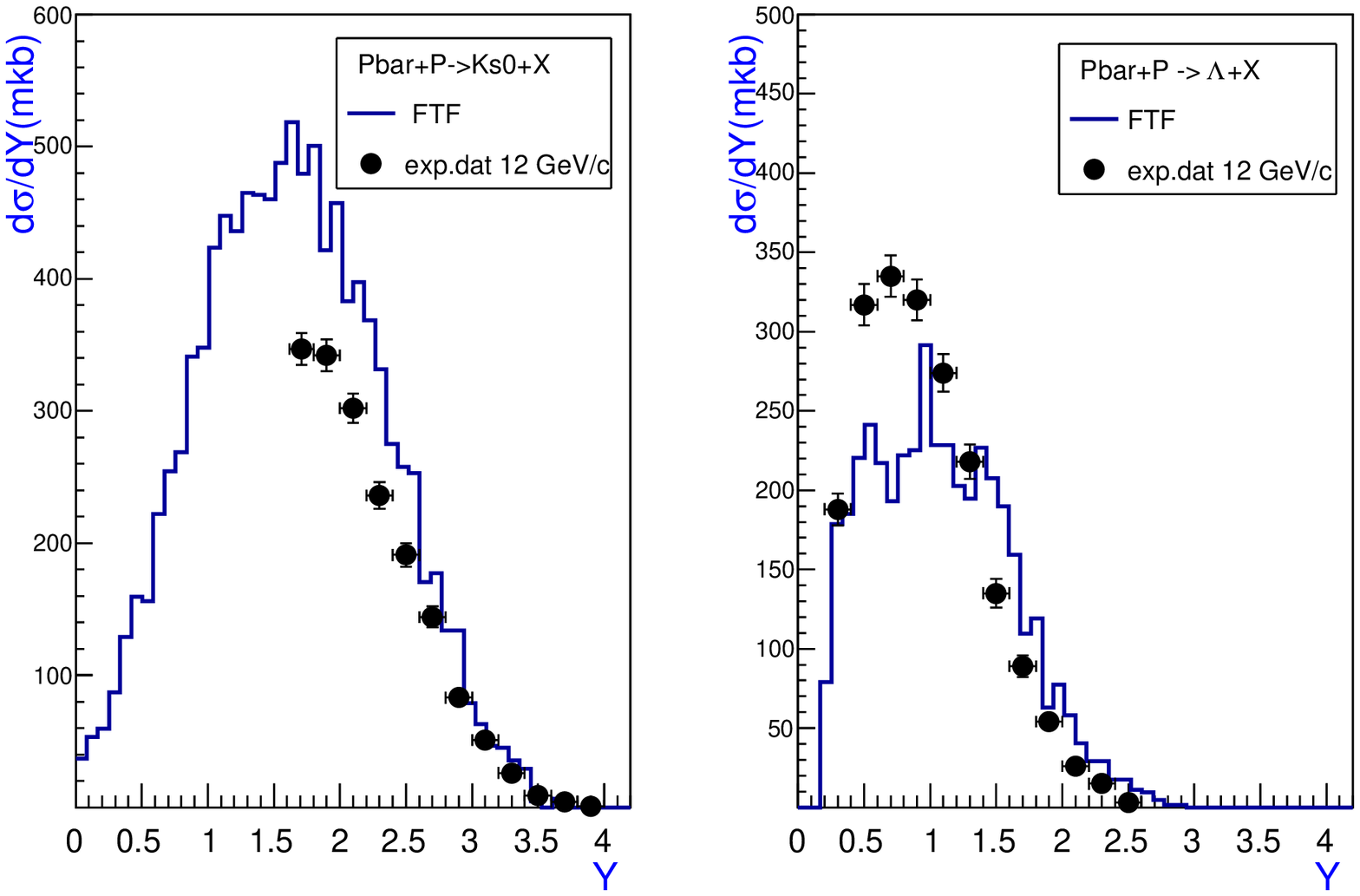}
\includegraphics[width=95mm,height=35mm,clip]{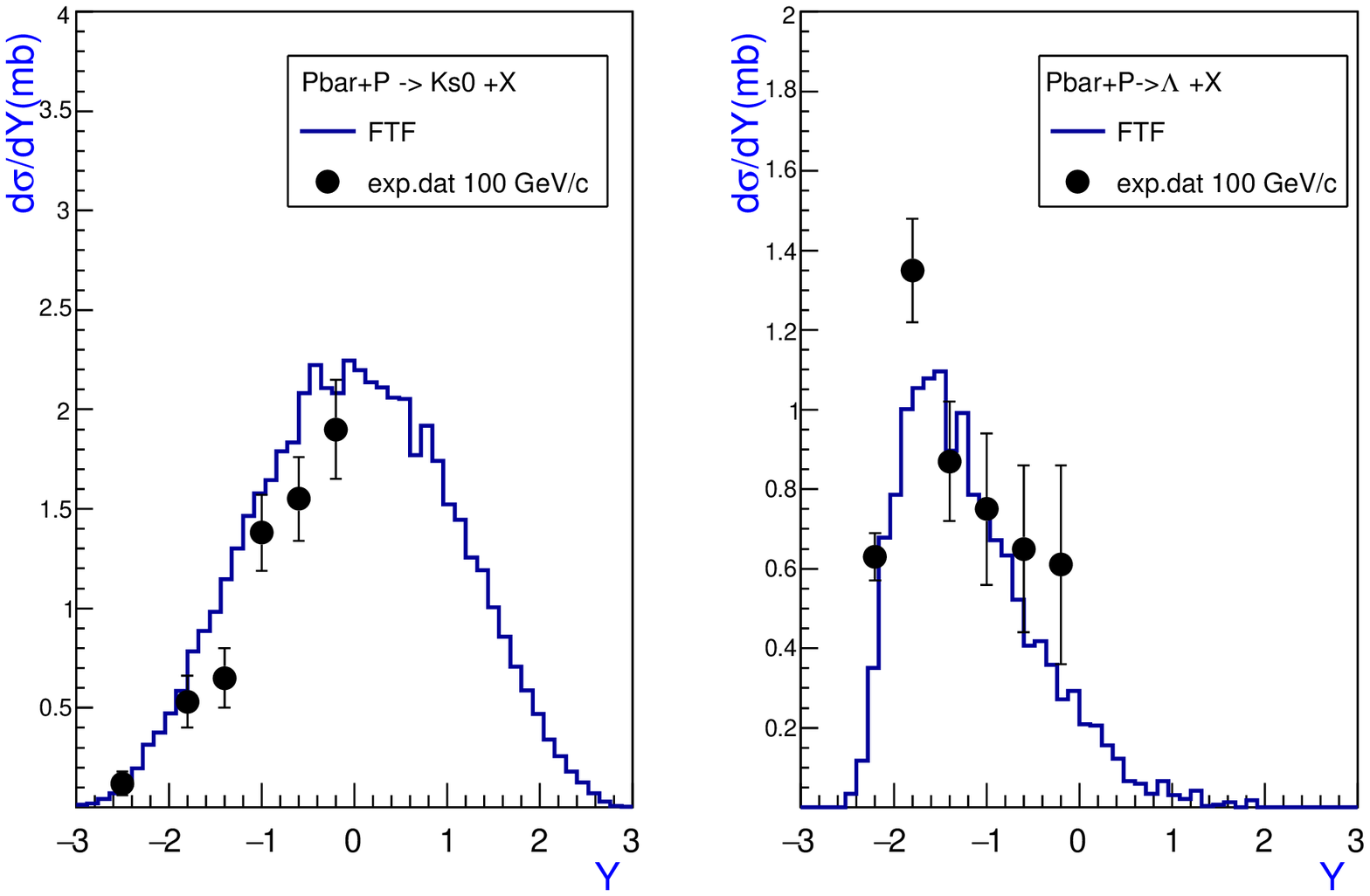}
  \caption{Rapidity distributions of $\Lambda$-hyperons and $K^0_S$ mesons at $P_{lab}=$ 
3.6, 12 and 100 GeV/c.
 The points are experimental data, the lines are the FTF model calculations.}
\end{center}
\label{Fig7}
\end{figure}

$\Lambda$ and $\bar \Lambda$ productions in target and
projectile fragmentation regions are connected with the
processes $\bar p + p\rightarrow \bar \Lambda + \Lambda$ at low
energies. At high energies, the processes can be $\bar p +
p\rightarrow \bar \Lambda + K^0 + p + m\ \pi$. As seen, $K^0$
mesons are mainly produced in the central region. It is natural
at high energies. At low energies,  $K^0$ mesons  can be
created in the processes $\bar p + p\rightarrow \bar \Lambda +
K^0 +n$ or $\bar p + p\rightarrow \bar p + K^0 + \bar K^0 + p$.

In Fig. 8, we present properties of $ \bar p +$Xe interactions
at 200 GeV/c. As seen, the FTF model describes general features
of strange particle production in $\bar p+A$ collisions. This
takes place also at lower energies.
\begin{figure}[cbth]
\begin{center}
\includegraphics[width=85mm,height=50mm,clip]{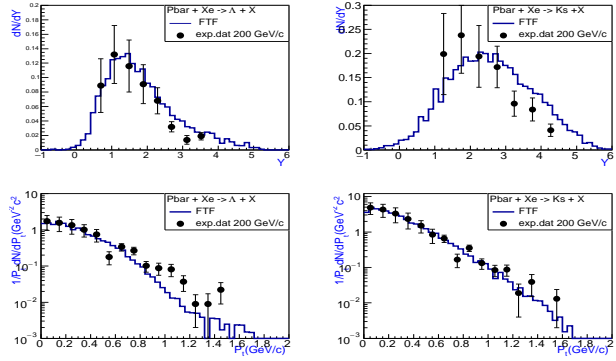}
\end{center}
  \caption{Rapidity and $P^2_T$ distributions of $\Lambda$-hyperons and $K^0_S$ mesons in $\bar p+Xe$
           interactions at 200 GeV/c. The points are experimental data, the lines are the FTF model calculations.}
\label{Fig8}
\end{figure}

\section{Conclusion}
The Dual Parton Model is implemented in the FTF model of Geant4 toolkit for
simulations of anti-proton-proton and anti-proton-nucleus interactions.
The model reproduces general features of the anti-proton interactions with nucleons
and nuclei starting from annihilation at rest up to $\sim$ 1000 GeV/c.

\section{Acknowledgements}
The authors are thankful to heterogeneous computing team of LIT
JINR   (HybriLIT) for support of our calculations.


\begin{thebibliography}{99}

\bibitem{Geant4}
S. Agostinelli ({\sc Geant4} Collaboration), \textit{Nucl. Instrum. Methods A} \textbf{506} (2003) 250;
J. Allison ({\sc Geant4} Collaboration), \textit{IEEE Trans. Nucl. Sci.}  \textbf{53} (2006) 270;
J. Allison ({\sc Geant4} Collaboration), \textit{Nucl. Instrum. Methods A} \textbf{835} (2016) 186.

\bibitem{DPM}
A.~Capella, U.~Sukhatme, C-I~Tan, J.~Tran~Thanh~Van, \textit{Phys. Rept.} \textbf{236} (1994) 225;
A.B.~Kaidalov and K.A.~Ter-Martirosian, \textit{Phys. Lett. B} \textbf{117} (1982) 247.

\bibitem{Glauber} R.J.~Glauber \textit{Lectures in Theoretical Physics} v. \textbf{1},
Intersci. Publishers, N.Y., 1959;

\bibitem{FrancoAA}V.Franco \textit{Phys. Rev.} \textbf{175} (1968) 1376;
\bibitem{CzyzAA}W.Czyz and L.C.Maximon \textit{Ann. of Phys. (N.Y.)} \textbf{52} (1969) 59.

\bibitem{Xs}V.V. Uzhinsky and A.S. Galoyan  \textit{arXive:0212369} [hep-ph] (2012).
\bibitem{Leap05}A.Galoian and V.Uzhinsky \textit{AIP Conf. Proc.} \textbf{796} (2005) 79.

\bibitem{Baldin}A. Galoian, A. Ribon and V.Uzhinsky \textit{PoS BaldinISHEPPXXII} (2015) 049.

\bibitem{Franco66}V. Franco and R.J. Glauber \textit{Phys. Rev.}  \textbf{142} (1966) 1195.
\bibitem{Dalk_Karm}O.D. Dalkarov and V.A. Karmanov \textit{Nucl. Phys. A} \textbf{445} (1985) 579.

\bibitem{OurPaperPL}V. Uzhinsky, J. Apostolakis, A. Galoyan et al. \textit{Phys. Lett.B}
                    \textbf{705} (2011) 235.

\bibitem{AGK}V.A. Abramovsky, V.N. Gribov and O.V. Kancheli \textit{Sov. J. Nucl. Phys.}
             \textbf{18} (1974) 308 (\textit{Yad. Fiz.} \textbf{18} (1973) 595).

\bibitem{AGaloyanSup} A. Galoyan and V.Uzhinsky
    \textit{Hyperfine Interact.} \textbf{215} (2013) 69.

\bibitem{BIC}G. Folger, V.N. Ivanchenko and J.P. Wellisch \textit{Eur. Phys. J. A}  \textbf{21} (2004) 407.

\bibitem{PbarAlow}
P.L.McGaughey et al. \textit{Phys. Rev. Lett.} \textbf{56} (1986) 2156;
H.J. Bersch et al. \textit{Zeit. fur Phys. A} \textbf{292} (1979) 197.

\end{thebibliography}
\end{document}